\documentclass[10pt,conference]{IEEEtran}
\IEEEoverridecommandlockouts
\usepackage[utf8]{inputenc}
\usepackage{amsmath,amssymb,amsfonts}
\usepackage{algorithmic}
\usepackage{graphicx}
\usepackage{textcomp}
\usepackage{xcolor}
\def\BibTeX{{\rm B\kern-.05em{\sc i\kern-.025em b}\kern-.08em
    T\kern-.1667em\lower.7ex\hbox{E}\kern-.125emX}}

\usepackage{booktabs,multirow}
\usepackage{subcaption} 

\usepackage{adjustbox}
\usepackage{tikz}
\usetikzlibrary{quantikz}
\usetikzlibrary{external}

\usepackage{amsthm}

\usepackage[colorlinks=true]{hyperref}

\usepackage[switch]{lineno}
\newcommand*\patchAMSlineno[1]{
	\expandafter\let\csname old#1\expandafter\endcsname\csname #1\endcsname
	\expandafter\let\csname oldend#1\expandafter\endcsname\csname end#1\endcsname
	\renewenvironment{#1}
		{\linenomath\csname old#1\endcsname}
		{\csname oldend#1\endcsname\endlinenomath}
}
\AtBeginDocument{
	\patchAMSlineno{equation}
	\patchAMSlineno{equation*}
	\patchAMSlineno{align}
	\patchAMSlineno{align*}
}

\NewDocumentCommand{\bbeta}{o}{
	\pmb{\beta}\IfValueT{#1}{_{[#1]}}
}
\NewDocumentCommand{\bgamma}{o}{
	\pmb{\gamma}\IfValueT{#1}{_{[#1]}}
}
\NewDocumentCommand{\bgstate}{o}{
	\ket{	\IfValueTF{#1}{\bbeta[#1]}{\bbeta},
		\IfValueTF{#1}{\bgamma[#1]}{\bgamma}
	}
}
\NewDocumentCommand{\bgstateT}{o}{
	\bra{	\IfValueTF{#1}{\bbeta[#1]}{\bbeta},
		\IfValueTF{#1}{\bgamma[#1]}{\bgamma}
	}
}

\begin{document}

\title{High-Round QAOA for MAX $k$-SAT on Trapped Ion NISQ Devices}
\author{
	\IEEEauthorblockN{Elijah Pelofske, Andreas Bärtschi, John Golden, Stephan Eidenbenz}
	\IEEEauthorblockA{\textit{CCS-3 Information Sciences} \\
	\textit{Los Alamos National Laboratory}\\
	Los Alamos, NM 87544, USA \\
	\{epelofske, baertschi, golden, eidenben\}@lanl.gov}
}

\maketitle
\thispagestyle{plain}
\pagestyle{plain}


\begin{abstract}
The Quantum Alternating Operator Ansatz (QAOA) is a hybrid classical-quantum algorithm that aims to sample the optimal solution(s) of discrete combinatorial optimization problems. We present optimized QAOA circuit constructions for sampling MAX $k$-SAT problems, specifically for $k=3$ and $k=4$. The novel $4$-SAT QAOA circuit construction we present uses measurement based uncomputation, followed by classical feed forward conditional operations. The QAOA circuit parameters for $3$-SAT are optimized via exact classical (noise-free) simulation, using HPC resources to simulate up to $20$ rounds on $10$ qubits. In order to explore the limits of current NISQ devices we execute these optimized QAOA circuits for random $3$-SAT test instances with clause-to-variable ratio $4$ on four trapped ion quantum computers: Quantinuum H1-1 (20 qubits), IonQ Harmony (11 qubits), IonQ Aria 1 (25 qubits), and IonQ Forte (30 qubits). The QAOA circuits that are executed include $n=10$ up to $p=20$, and $n=22$ for $p=1$ and $p=2$. The high round circuits use upwards of 9,000 individual gate instructions, making these some of the largest QAOA circuits executed on NISQ devices. Our main finding is that current NISQ devices perform best at low round counts (i.e., $p = 1,\ldots, 5$) and then -- as expected due to noise -- gradually start returning satisfiability truth assignments that are no better than randomly picked solutions as the number of QAOA rounds are further increased. 
\end{abstract}


\section{Introduction}
\label{section:intro}

The Quantum Approximate Optimization Algorithm (QAOA) \cite{Farhi2014, farhi2015quantum}, which has subsequently been generalized to the Quantum Alternating Operator Ansatz \cite{Hadfield2019} to include a wider variety of mixers and state preparation algorithms \cite{nasa2020XY, baertschi2020grover, Golden_2021, golden2022evidence, chancellor2019domain, larose2022mixer, he2023alignment}, under the same acronym of QAOA, is a quantum algorithm that is intended to sample optimal solutions of discrete combinatorial optimization problems. 

QAOA has been applied to a number of types of combinatorial optimization problems, both numerically and experimentally on near term quantum computers, including on boolean $k$-SAT problems \cite{Zhang_2022, boulebnane2022solving, golden2023quantum}, maximum cut \cite{https://doi.org/10.48550/arxiv.2303.02064, herrman2021impact, harrigan2021quantum, obst2023comparing}, and the Sherrington-Kirkpatrick model \cite{harrigan2021quantum, Farhi_2022}. How QAOA scales with respect to problem size and number of rounds \cite{Weidenfeller2022scalingofquantum, PhysRevX.10.021067}, how learn-able the optimal (or near optimal) QAOA parameters are in the presence of noise \cite{Wang_2021}, and many other related questions are topics of interest for investigating problem-specific QAOA settings for potential advantageous quantum computation \cite{farhi2019quantum, shaydulin2023evidence}. The primary challenge is managing the relatively high error rates on current hardware. 

While the suitability of QAOA for the NISQ era~\cite{Preskill2018quantumcomputingin} is thus an ongoing debate, particularly because in general QAOA requires high $p$ in order to be effective \cite{farhi2020quantum, farhi2020quantum_2}, we instead propose to study QAOA performance as a benchmark-style test for NISQ devices, where we expect solutions to initially improve as we increase the rounds of QAOA before hitting a peak and then degrading to random solutions. The higher number of QAOA rounds a NISQ device can maintain or even increase the average quality of solutions sampled, the better its performance. 

\begin{figure}[t!]
    \centering
    \includegraphics[width=\linewidth]{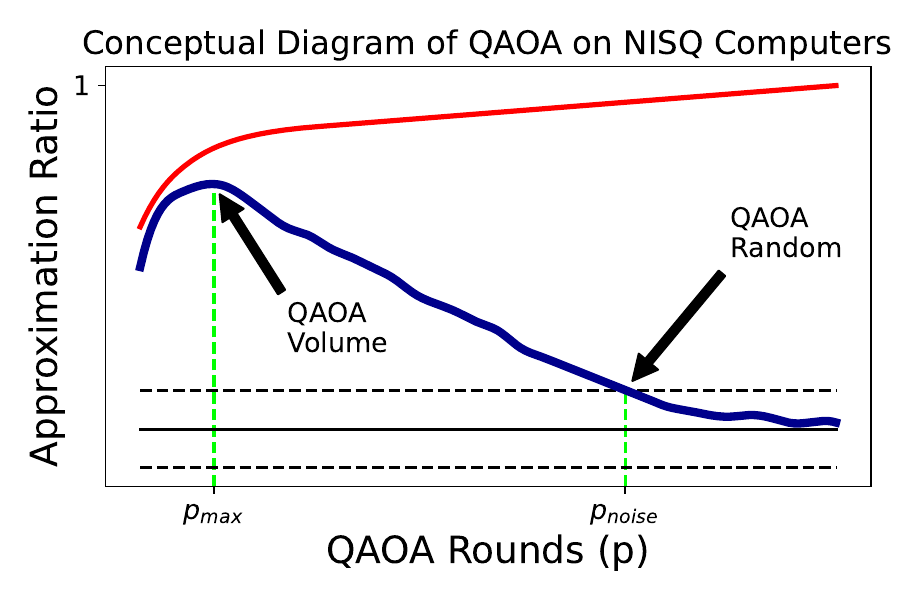}
    \caption{Conceptual diagram of using QAOA to characterize the performance of NISQ devices for general combinatorial optimization problems. The horizontal black lines show the random sampling distribution, the red line shows the ideal (zero noise) quantum computation, and the blue line is a conceptual sketch of the expected QAOA performance when noise is present in the computation. The points $p_{\max}$ and $p_{\mathrm{noise}}$ are marked with vertical dashed green lines. }
    \label{fig:conceptual_diagram}
\end{figure}

{\it NISQ QAOA Performance Measures.} To be more precise and as illustrated conceptually in Figure~\ref{fig:conceptual_diagram}, we define the quality achieved in a QAOA NISQ experiment as the average approximation ratio of all solutions sampled by the device, where the approximation ratio of a solution $s$ with objective function value $C(s)$ is defined as $\frac{C(s)}{C_{opt}}$ (for a maximization problem, and inverted for a minimization problem) with $C_{opt}$ being the optimal objective function value across all feasible solutions. This is under the assumption that good angles, but not necessarily optimal angles, can be learned efficiently for each round $p$; in other words the conceptual picture of NISQ QAOA in Figure \ref{fig:conceptual_diagram} is not intended to capture the difficulty of the angle finding problem (especially at very high rounds). We plot approximation ratios on the $y$-axis, and put the length of the QAOA calculation, crudely measured in number of rounds $p$ on the $x$-axis. On a noise-free or error-corrected quantum computing device (see red line), we expect the approximation ratio to gradually increase  with $p$ until it eventually hits 1.0 when QAOA has put all amplitude into the optimum solution(s). On the other hand, if we have a device that samples purely randomly because of its overwhelming noise level, it would sample random solutions perhaps mostly within a few percentile points above the average random solution (marked by the dashed horizontal lines in Figure \ref{fig:conceptual_diagram}). We expect a NISQ device (dark blue line) to initially improve approximation ratios to reach a peak value at the point labeled QAOA Volume at $p_{\max}$, which is a measure defined as $n \times p$, where $n$ is the number of qubits used. As we further increase $p$ beyond the peak, we expect noise to gradually become more dominant until the solution quality becomes indistinguishable from random samples and we have thus lost all (quantum) signal. Let us define a point at $p_{{noise}}$, where the NISQ performance reaches the upper envelope of the 60th percentile of  standard deviation from random sampling, where we have chosen the number 60 in an arbitrary but fixed fashion. In particular, the motivation for this tight percentile choice is based on the sampling characteristics of the $3$-SAT problems that we will be sampling, for other combinatorial optimization problems it may make sense to use other reasonable percentiles. We call this point ``QAOA Random''. The pair $(p_{\max}, p_{\mathrm{noise}})$ characterizes QAOA performance on a NISQ device and it is of particular interest to study these components as we vary problem instance sizes. 

For an error-corrected device, we would define $p_{\max}$ to be the smallest $p$ that reaches an approximation ratio that we wish to achieve, ideally we would hope that it is small and grows at most polynomially with the problem instance size as it becomes algorithmically defined by QAOA performance. In contrast, on a noisy device, we hope for $p_{\max}$ to be as large as possible as it serves much more as an indicator that the NISQ device still improves signal at high round counts. The QAOA Random point with $p_{\mathrm{noise}}$ would be at infinity for an error-corrected device. In the reality of NISQ devices, we hope for $p_{\mathrm{noise}}$ to be as large as possible.

In this paper, we test how well NISQ devices match our proposed framework of QAOA performance. We choose the well-known MAX $3$-SAT problem as our object of study. We are specifically interested in observing what happens to the QAOA computation on NISQ  computers when pushed to very high rounds, consisting of several thousand single and two qubit gates. To this end, we experimentally utilize four trapped ion based quantum computers, IonQ Harmony \cite{Wright_2019}, IonQ Aria, IonQ Forte \cite{chen2023benchmarking}, and Quantinuum H1-1 \cite{Pino_2021} in order to push these QAOA computations to very high rounds (up to $p=20$), where each round has a high gate depth (approximate gate depth of $200$ per round). 

{\it Optimized k-SAT QAOA Circuits. } We present optimized $k$-SAT transverse field QAOA circuit construction algorithms, in particular for $3$-, and $4$-SAT, targeting all-to-all connectivities with native $\texttt{Rzz}$ two-qubit gates. The $4$-SAT implementation makes use of mid-circuit measurement based AND uncomputation with mid circuit qubit resets in order to re-use ancilla qubits for the computation. Therefore, implementing the $4$-SAT QAOA circuits can additionally provide a future benchmark on the capability of quantum computers combined with classical co-processing (and mid circuit measurement) for feed forward conditioning. The $4$-SAT implementation requires even more gate operations and gate depth per round compared to $3$-SAT, and puts it well out of reach of the state-of-the-art quantum computers at this moment. However, we believe that such $4$-SAT QAOA circuits may be within reach for new quantum computers in the coming years, and therefore we provide these circuit construction algorithms as a benchmark for future quantum computers, with the intent being that very high round QAOA will measure the computational reach for very high depth quantum circuits. 

Formally, in the problem of Maximum $k$-Satisfiability (or Max $k$-SAT), we are given a logical formula in Conjunctive Normal Form (CNF) defined on $n$ binary truth variables in the variable vector $x := x_0, \ldots, x_{n-1}$ with $x_j = \{0,1\}$  for $j= 0, \ldots, n-1$. The CNF formula can be expressed as a conjunction of disjunctions (more commonly known as clauses), where each clause is defined on $k$ literals, which are instances or negated instances of the variables. 
Furthermore we require each variable to appear at most once per clause.
The goal is to find a truth value assignment for the variables such that a maximum number of clauses are satisfied; thus, we define an objective function $C(x)\colon \{1,0\}^n \to \mathbb{N}$, where $C(x)$ is the number of clauses satisfied by truth assignment $x$. Max $k$-SAT is $NP$-hard for $k \geq 2$, and no classical polynomial-time approximation algorithm can guarantee to find a solution that is within $\frac{2^k-1}{2^k} + \epsilon$ of the optimum \cite{haastad2001some} (for any arbitrary but fixed $\epsilon >0$); note that the decision version of 2-SAT can be solved classically polynomially. Consider an example of MAX $3$-SAT with $n=3$ (though in general $n\neq k$): 
\begin{align*}
C(x) =&\ (\overline{x_0} \vee x_1 \vee x_2) + (\overline{x_0} \vee \overline{x_1} \vee x_2)  +  \\
&\ (\overline{x_0} \vee \overline{x_1} \vee  \overline{x_2}) + (x_0 \vee x_1 \vee x_2)  \nonumber 
\end{align*}
We have liberally mixed logical notation (e.g., the $\vee$ for logical OR) and arithmetic notation to best serve our purpose. 
The first clause,\footnote{which we will also use to illustrate the circuit design in Section~\ref{section:methods_k_SAT_circuits},} 
$(\overline{x_0} \vee x_1 \vee x_2)$, is satisfiable (i.e. equals~1), when at least one of the literals $\overline{x_0}, x_1, x_2$
evaluates to \texttt{True}. Equivalently, the only unsatisfiable assignment is $(x_0,x_1,x_2) \leftarrow (\texttt{True},\texttt{False},\texttt{False}) = (1,0,0)$.
Thus, of the $2^3 = 8$ possible truth assignments, the four assignments $(0,0,1), (0,1,0), (0,1,1), (1,0,1)$ result in all four clauses being satisfied, e.g. $C((0,1,0)) = 4$, while $(1,0,0), (1,1,0), (0,0,0), (1,1,1)$ result in only three satisfied clauses (where each solution vector is $(x_0, x_1, x_2))$.

{\it Related work.} 
With respect to qubit count, QAOA has been applied to short depth QAOA circuits on IBMQ hardware lattices up to 127 qubits \cite{pelofske2023quantum}. However, with respect to circuit depth and overall gate count, the results we present here are significantly larger than previous studies. 
QAOA has been implemented on Quantinuum quantum computers in several studies \cite{https://doi.org/10.48550/arxiv.2303.02064, niroula2022constrained, https://doi.org/10.48550/arxiv.2206.03144, Lotshaw_2022}, however the high round and large problem size QAOA circuits evaluated in this paper are the largest QAOA experimental implementations yet presented with respect to gate counts and depths of the circuits that the authors are aware of. We note in particular, that the impressive work \cite{https://doi.org/10.48550/arxiv.2303.02064} on MaxCut on 3-regular graphs on Quantinuum H1-1 device achieves a higher QAOA volume than our work, where QAOA volume is defined as $n \cdot p$ with $n$ being the number of qubits; however, the phase separator for MaxCut on 3 regular graphs consist of $1.5n$ two-qubit gates (or $3n$ qubit gates if only using CNOT gates), whereas our 3-SAT phase separator requires $20n$ two-qubit gates (details in later sections).

There are a growing number of QAOA variants, for example using different mixers and phase separators \cite{nasa2020XY, baertschi2020grover, Golden_2021, golden2022evidence, larose2022mixer, he2023alignment, Hadfield_2022, zhu2022multiround}, however for the circuit construction algorithm and NISQ computer experiments we present, the standard transverse field mixer is used.

\section{Methods}
\label{section:methods}
In Section~\ref{section:methods_QAOA_introduction}, we briefly overview QAOA. Section \ref{section:methods_k_SAT_circuits} details a general algorithm for constructing QAOA circuits for $k$-SAT problems. Section \ref{section:methods_hardware_implementation} describes the hardware implementations on IonQ Harmony and Quantinuum H1-1.

\begin{figure}[t!]
	\centering
	\newcommand{\rxgate}[1]{\gate{R_x(2\beta_{#1})}}
	\newcommand{\rygate}{\gate{H}}
	\begin{adjustbox}{width=\linewidth}
	\begin{quantikz}[row sep={24pt,between origins},execute at end picture={
				\node[xshift=30pt, yshift=-24pt] at (\tikzcdmatrixname-5-4) {
					$\underbrace{\hspace*{220pt}}_{p\text{ rounds with angles } \gamma_1,\beta_1,\ldots,\gamma_p,\beta_p}$
				};
			}]
		\lstick{\ket{0}}	
		& \rygate	
		& \gate[5]{e^{-i\gamma_1 H_C}}
		& \rxgate{1}
		& \qw\midstick[5,brackets=none]{\ \ldots\ }
		& \rxgate{p}
		& \meter{}\rstick[5]{\rotatebox{90}{\large $\bgstateT H_C \bgstate$}}
		\\	
		\lstick{\ket{0}}	& \rygate	&	& \rxgate{1}	&\qw	& \rxgate{p}	& \meter{}		\\	
		\lstick{\ket{0}}	& \rygate	&	& \rxgate{1}	&\qw	& \rxgate{p}	& \meter{}		\\	
		\lstick{\ket{0}}	& \rygate	&	& \rxgate{1}	&\qw	& \rxgate{p}	& \meter{}		\\	
		\lstick{\ket{0}}	& \rygate	&	& \rxgate{1}	&\qw	& \rxgate{p}	& \meter{}	
	\end{quantikz}
	\end{adjustbox}
	\caption{Main QAOA subroutine: State Preparation, $p$ rounds of Phase Separator and Mixer applications, Measurement.\linebreak
	Both state preparation and each Transverse Field mixer can be implemented with single-qubit gates,
	$H\ket{0} = \ket{+}$ 
    and $R_x(2\beta) = e^{-i\beta X}$.%
    }
    \label{fig:qaoa-circuit}%
\end{figure}
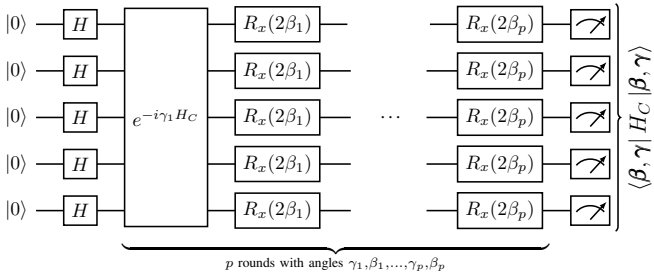

\subsection{QAOA overview}
\label{section:methods_QAOA_introduction}

Given an discrete variable assignment combinatorial optimization problem, we can define a an objective function of the form $C(x)\colon \{1,0\}^n \to \mathbb{N}$ which provides a real number evaluation for any variable assignments. In this case, where the combinatorial optimization problem is maximization, we aim to find variable assignments that provide the maximum possible objective function evaluation over all possible inputs. The QAOA algorithm intends to sample a combinatorial optimization of this form. The ingredients of QAOA are as follows:
\begin{itemize}
    \item an initial state $\ket{\psi}$, here the equal superposition over all computational basis states $\ket{\psi}=\ket{+^n}$,
    \item a classical \texttt{phase separating} Cost Hamiltonian $H_C = \sum_{x\in \{0,1\}^n} C(x)\ket{x}\bra{x}$,
    \item a \texttt{mixing} Hamiltonian, here the Pauli-X transverse field $H_M = \sum_{i=1}^n X_i$,
    \item an integer $p\geq 1$, the number of subroutine rounds (also referred to as \emph{layers}),
    \item two vectors composed of real numbers $\bgamma = (\gamma_1,...,\gamma_p)$ and $\bbeta = (\beta_1,...,\beta_p)$, each vector having a length of $p$.
\end{itemize}
The quantum subroutine of QAOA, depicted in Figure~\ref{fig:qaoa-circuit}, then prepares a state 
\begin{equation}
    \bgstate = e^{-i\beta_p H_M} e^{-i\gamma_p H_C} \cdots e^{-i\beta_1 H_M} e^{-i\gamma_1 H_C} \ket{\psi}
    \label{eq:bgstate}
\end{equation}
and samples solutions from it, with an expectation value of $\bgstateT H_C \bgstate$.
To utilize QAOA, the classical outer loop routine is to find these real vectors $\vec{\gamma}$ and $\vec{\beta}$, typically referred to as the \emph{angles}, such that the expectation value is maximized (or minimized). Importantly at low rounds (e.g. $p=1, 2$) it is not expected that we will be able to reliably sample the optimal solutions, especially for very large problems. In order to increase the probability of sampling good solutions, we must increase the number of rounds; as we increase the number of rounds, assuming we are able to learn good angles at each round, we expect a consistent increase of the ideal computation of the approximation ratio $\bgstateT H_C \bgstate / C_{\mathrm{opt}}$
of the sampled problem. QAOA is considered a type of variational quantum algorithm \cite{Cerezo_2021}, or a hybrid quantum-classical algorithm, because in the absence of knowing good angles $\beta$ and $\gamma$ a-priori, it is likely the case that the computation would need evaluate the quality of some angle combination using a quantum computer, and then update those angles using machine learning on a classical co-processor. Errors in the quantum computation make this angle finding task considerably more difficult however, and in the case of the experiments we present, we classical learn high quality angles to use for the QAOA computation on quantum computer. In this respect, this study is not aiming to determine the learn-ability of high round QAOA angles, but more so the ideal and the NISQ QAOA scaling of approximation ratios with respect to problem size and increasing rounds.

\begin{figure*}[t!]
	\centering
    \newcommand{\ORgate}{\gate[3]{\text{\Large $\ e^{-i\gamma\left(\overline{x_0}\, \vee\, x_1\, \vee\, x_2\right)}\ $}}}
	\newcommand{\ANDgate}{\gate[3]{\text{\Large $\ e^{-i\gamma\left(1 - (x_0\, \wedge\, \overline{x_1}\, \wedge\, \overline{x_2})\right)}\ $}}}
	\newcommand{\rzgate}[3]{\gate{R_z(#1\tfrac{#2}{#3})}}
	\newcommand{\rzzgate}[3]{\gate[2]{R_{zz}(#1\tfrac{#2}{#3})}}
	\begin{adjustbox}{width=\linewidth}
	\begin{quantikz}[row sep={24pt,between origins},execute at end picture={}]
		\lstick{$q_0\colon$}	& \ORgate	& \qw	& 	& & \ANDgate	& \qw	& 	& & \qw		& \phase{\gamma}	& \qw		& \qw	& 	& & \qw		& \rzgate{+}{\gamma}{4}	 & \ctrl{2}	& \qw			& \qw		& \qw			& \ctrl{2}	& \rzzgate{-}{\gamma}{4}& \qw		& \qw   & \qw		\\
		\lstick{$q_1\colon$}	& \qw		& \qw	& =	& & 		& \qw	& \cong	& & \gate{X}	& \ctrl{-1}		& \gate{X}	& \qw	& =	& & \gate{X}	& \rzgate{+}{\gamma}{4}	 & \qw		& \qw			& \ctrl{1}	& \qw			& \qw		& 			& \ctrl{1}	& \gate{X}    & \qw	\\
		\lstick{$q_2\colon$}	& \qw		& \qw	& 	& & 		& \qw	& 	& & \gate{X}	& \ctrl{-1}       	& \gate{X}	& \qw	& 	& & \gate{X}	& \rzgate{+}{\gamma}{4}	 & \targ{}	& \rzgate{-}{\gamma}{4}	& \targ{}	& \rzgate{+}{\gamma}{4}	& \targ{}	& \rzgate{-}{\gamma}{4}	& \targ{}	& \gate{X}  & \qw
	\end{quantikz}
	\end{adjustbox}
	\caption{Phase Separator for a single $3$-SAT clause: The logical OR of three binary variables corresponds to the negated AND of the negated variables. 
	We neglect the global phase of $e^{-i\gamma \cdot 1}$ and phase the only non-satisfiable assignment $\ket{100}$ by $e^{+i\gamma}$. 
    The 2-controlled phase gate $\mathrm{CCPhase}(\gamma) = \mathrm{diag}(1,1,1,1,1,1,1,e^{i\gamma})$ we decompose with 5 two-qubit gates (4 CNOT, 1 \texttt{Rzz}), analogous to standard Toffoli decompositions~\cite{Barenco1995,schuch2003programmable}, which yields the optimal two-qubit gate count~\cite{yu2013five}.
	}
	\label{fig:PS-clause}
\end{figure*}
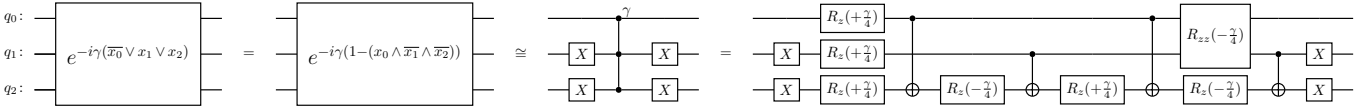

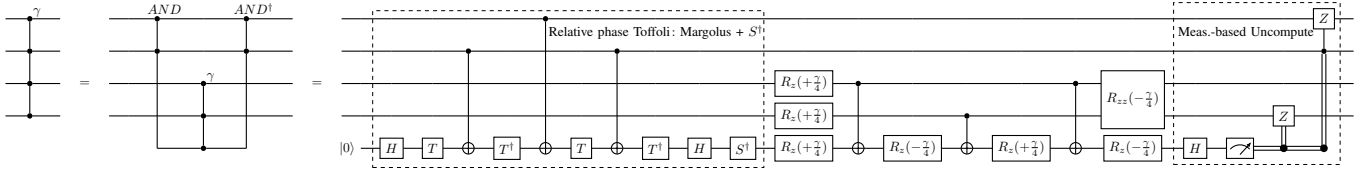
\begin{figure*}[t!]
	\centering
	\newcommand{\rzgate}[3]{\gate{R_z(#1\tfrac{#2}{#3})}}
	\newcommand{\rzzgate}[3]{\gate[2]{R_{zz}(#1\tfrac{#2}{#3})}}
	\begin{adjustbox}{width=\linewidth}
	\begin{quantikz}[row sep={24pt,between origins},execute at end picture={
				\node[fit=(\tikzcdmatrixname-1-14)(\tikzcdmatrixname-5-23),draw,dashed,inner ysep=6pt,inner xsep=10pt,xshift=-4pt,label={[yshift=-25pt,xshift=65pt]Relative phase Toffoli$\colon$Margolus + $S^{\dagger}$}] {};				
				\node[fit=(\tikzcdmatrixname-1-31)(\tikzcdmatrixname-5-34),draw,dashed,inner ysep=12pt,inner xsep=14pt,xshift=-2pt,label={[yshift=-31pt,xshift=-8pt]Meas.-based Uncompute}] {};				
			}]
		\qw	& \phase{\gamma}	& \qw	& 	& & \qw		& \phase{AND}		& \qw		& \phase{AND^{\dagger}}	& \qw	& 	& & \qw			& \qw		& \qw		& \qw		& \qw			& \ctrl{4}	& \qw		& \qw		& \qw			& \qw		& \qw			& \qw				 & \qw		& \qw			& \qw		& \qw			& \qw		& \qw			& \qw		& \qw		& \qw		& \gate{Z}	& \qw	\\
		\qw	& \ctrl{-1}		& \qw	& 	& & \qw		& \ctrl{3}\vqw{-1}	& \qw		& \ctrl{3}\vqw{-1}	& \qw	& 	& & \qw			& \qw		& \qw		& \ctrl{3}	& \qw			& \qw		& \qw		& \ctrl{3}	& \qw			& \qw		& \qw			& \qw				 & \qw		& \qw			& \qw		& \qw			& \qw		& \qw			& \qw		& \qw		& \qw		& \ctrl{-1}	& \qw	\\
		\qw	& \ctrl{-1}		& \qw	& =	& & \qw		& \qw			& \phase{\gamma}& \qw			& \qw	& =	& & \qw			& \qw		& \qw		& \qw		& \qw			& \qw		& \qw		& \qw		& \qw			& \qw		& \qw			& \rzgate{+}{\gamma}{4}	 & \ctrl{2}	& \qw			& \qw		& \qw			& \ctrl{2}	& \rzzgate{-}{\gamma}{4}& \qw		& \qw		& \qw		& \qw		& \qw	\\
		\qw	& \ctrl{-1}		& \qw	& 	& & \qw		& \qw			& \ctrl{-1}	& \qw			& \qw	& 	& & \qw			& \qw		& \qw		& \qw		& \qw			& \qw		& \qw		& \qw		& \qw			& \qw		& \qw			& \rzgate{+}{\gamma}{4}	 & \qw		& \qw			& \ctrl{1}	& \qw			& \qw		& 			& \qw		& \qw		& \gate{Z}	& \qw		& \qw	\\
		\	& 		    	& 	& 	& & 		& 			& \ctrl{-1}	& \qw			& 	& 	& & \lstick{\ket{0}}	& \gate{H}	& \gate{T}	& \targ{}	& \gate{T^{\dagger}}	& \targ{}	& \gate{T}	& \targ{}	& \gate{T^{\dagger}}	& \gate{H}	& \gate{S^{\dagger}}	& \rzgate{+}{\gamma}{4}	 & \targ{}	& \rzgate{-}{\gamma}{4}	& \targ{}	& \rzgate{+}{\gamma}{4}	& \targ{}	& \rzgate{-}{\gamma}{4}	& \gate{H}	& \meter{}	& \cwbend{-1}	& \cwbend{-3}	&		
	\end{quantikz}
	\end{adjustbox}
	\caption{Phase Separator for a single 4-SAT clause (focusing on negative literals): The number of quantum controls on the phase shift gate can be decreased by a pairwise AND computation into an ancilla qubit, followed by the smaller controlled phase shift and an AND$^{\dagger}$ uncomputation.
    The AND can be implemented with only 3 CNOTs using a relative phase Toffoli implementation~\cite{maslov2016advantages, song2003simplified} given by Qiskit's Margolus gate RCCX and a consecutive $S^{\dagger}$.\\
    The AND$^{\dagger}$ can be implemented with measurement based uncomputation~\cite{Gidney2019}, utilizing mid circuit measurement (followed by qubit reset) and a classical feed forward condition for a single CZ two-qubit gate. Furthermore, this also allows to replace the last CNOT of the 2-controlled phase shift from Figure~\ref{fig:PS-clause} to be moved to a classically controlled single-qubit gate.\\
    Thus we used $3$ CNOTs and $1$ classically controlled CZ to intermittently decrease the number of quantum controls on the phase shift by one, 
    plus another $3$ CNOTs and $1$ \texttt{Rzz} Gate to implement the phase shift, for a total of 8 two-qubit gates. For a $k$-SAT clause with $k>3$, recursively nesting this construction generalizes to a decomposition using $k-3$ ancilla qubits and $4\cdot (k-3) + 4 = 4k-8$ two-qubit gates.
    }
	\label{fig:PS-4SAT}
\end{figure*}

\subsection{QAOA Circuit Construction for MAX k-SAT}
\label{section:methods_k_SAT_circuits}

To design a circuit that prepares and samples from the state $\bgstate$ in Equation~\eqref{eq:bgstate}, we need to take care of the state preparation of $\ket{+^n}$, mixers of the form $e^{-i\beta \sum_i X_i}$, and phase separators of the form $e^{-i\gamma H_C}$. As illustrated in Figure~\ref{fig:qaoa-circuit}, the first two are achieved with simple one-qubit gates. 

To implement the phase separator for MAX $k$-SAT, we write first write the cost function $C(x)$ as a sum of $m$ clause functions $C_j(x)\colon \{0,1\}^n\rightarrow \{0,1\}$, each selecting $k$ \emph{distinct} variables $x_{j1},\ldots,x_{jk}$ variables of the input with positive or negative literal
(i.e. either $\ell_{ji}(x_{ji})=x_{ji}$ or $\ell_{ji}(x_{ji}) = \overline{x_{ji}}$):
\begin{align*}
    C(x)    & = \sum_{j=1}^m C_j(x)         \\
    C_j(x)  & = \left( \ell_{j1}(x_{j1})\ \vee \ \ell_{j2}(x_{j2})\ \vee \ \ldots \ell_{jk}(x_{jk}) \right)
\end{align*}

We can then rewrite the cost Hamiltonian $H_C$ as 
\begin{align*}
    H_C 
        & = \quad \sum_{\mathclap{x\in\{0,1\}^n}} \quad \sum_{j=1}^m \quad C_j(x) \ket{x}\bra{x} \\
        & = \quad \sum_{j=1}^m \underbrace{\quad\sum_{\mathclap{x\in\{0,1\}^n}} \quad C_j(x) \ket{x}\bra{x}}_{=: H_{C_j}}.
\end{align*}

Since $H_C$ is diagonal in the computational basis and the terms commute, we can implement $e^{-i\gamma H_C}$ as $\prod_{j=1}^m e^{-i\gamma H_{C_j}}$,
meaning we implement the phasing for each clause individually. Furthermore, only $k$ variables appear in each clause, making $H_{C_j}$ $k$-local. 
We now describe efficient circuits for $k=3,4$ in detail using Figures~\ref{fig:PS-clause} and~\ref{fig:PS-4SAT}. We note that our construction easily extends to arbitrary $k$-SAT clauses. 

\paragraph*{Clause Phase Separators}
Our goal is to compile $e^{-i\gamma H_{C_j}}$ efficiently with the respect to our NISQ hardware. In particular, we assume all-to-all connectivity and native parameterizable $\texttt{Rzz}(2\gamma) = e^{-i\gamma Z\otimes Z}$ two-qubit gates. The main cost in both execution time and source of error on NISQ devices is the number of two-qubit gates, hence we aim to minimize their use per clause. A single CNOT gate can always be implemented (up to single-qubit gates) with a single \texttt{Rzz} gate, but \emph{not vice versa} (depending on the angle $\gamma$). Hence our approach utilizes both CNOT and \texttt{Rzz} where applicable.

For a $k$-SAT clause $C_j$, note that applying $e^{-i\gamma H_{C_j}}$ amounts to phase all $2^k-1$ satisfiable variable assignments by $e^{-i \gamma}$. 
This is---up to a global phase of $e^{-i\gamma}$---equivalent to instead phasing the \emph{only} unsatisfiable assignment by $e^{+ i\gamma}$. We illustrate this for 
$C_j(x) = \left(\overline{x_0}\, \vee\, x_1\, \vee\, x_2\right)$, see Figure~\ref{fig:PS-clause}:
\begin{align*}
    e^{-i\gamma H_{C_j}}\ket{x_0x_1x_2} = & 
    \begin{cases}
        e^{-i\gamma} \ket{x_0x_1x_2}    & \text{if $x \neq (1,0,0$)} \\
        1 \ket{x_0x_1x_2}               & \text{otherwise},
    \end{cases}\\
    = e^{-i\gamma}\cdot &
    \begin{cases}
        e^{+i\gamma} \ket{x_0x_1x_2}    & \text{if $x = (1,0,0$)} \\
        1 \ket{x_0x_1x_2}               & \text{otherwise}.
    \end{cases}
\end{align*}
Evaluating whether the clause $C_j$ is unsatisfied amounts to evaluate whether the AND of its negated literals is satisfied, 
$1-\left(\overline{x_0}\, \vee\, x_1\, \vee\, x_2\right) = \left(x_0\, \wedge\, \overline{x_1}\, \wedge\, \overline{x_2}\right)$. 
The latter is true if and only if 
(i) \emph{{\bfseries after negating}} variables $x_1$ and $x_2$ (which in $C_j$ have positive literals), we have
(ii) $\left(x_0\, \wedge\, x_1\, \wedge\, x_2\right) =$ \texttt{True}.

\begin{figure*}[t!]
    \centering
    \includegraphics[width=\textwidth]{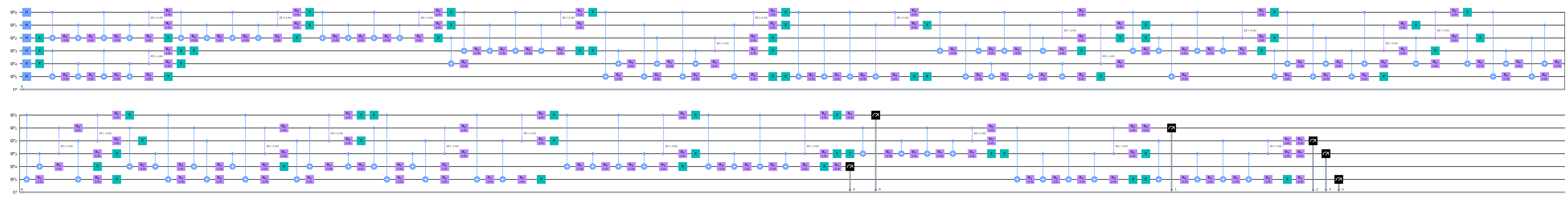}
    \caption{Example QAOA circuit for $p=1$ on a single random $n=6$ $3$-SAT instance, with clause density of $4$. This circuit is defined in terms of the gateset \texttt{h, x, cx, rzz, rz}. }
    \label{fig:non_compiled_circuit}
\end{figure*}

\begin{figure*}[t!]
    \centering
    \includegraphics[width=\textwidth]{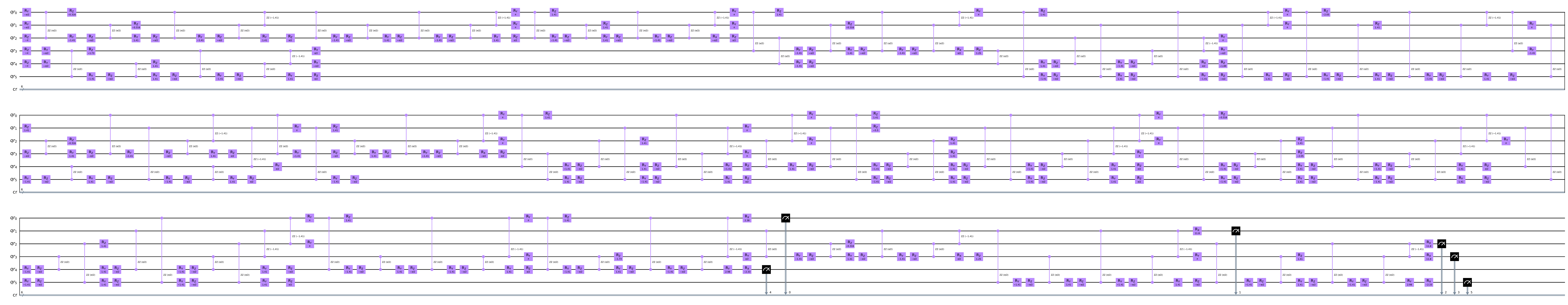}
    \caption{Compiled quantum circuit drawing for $p=1$ on a single random $n=6$ $3$-SAT instance, with clause density of $4$. Compiled connectivity is all-to-all, and the target gateset is \texttt{rzz, rz, ry, rz}. This circuit is the (Qiskit) compiled form of Figure \ref{fig:non_compiled_circuit}. }
    \label{fig:compiled_circuit}
\end{figure*}

\paragraph*{Circuit Examples and Gate Counts}
Implementing (i) is done by simply conjugating (ii) with $X$-gates on qubits $q_1$ and $q_2$. 
In Figure~\ref{fig:PS-clause}, we then show how to implement the $3$-SAT clause phase separator $e^{-i \gamma \left(x_0\, \wedge\, x_1\, \wedge\, x_2\right)}$. 
This operation corresponds to a 2-controlled $\mathrm{CCPhase}(\gamma) = \mathrm{diag}(1,1,1,1,1,1,1,e^{i\gamma})$ gate, which we decompose using 4 CNOTs and 1 \texttt{Rzz} gate, analogous to standard Toffoli decompositions~\cite{Barenco1995,schuch2003programmable}. Since 5 two-qubit gates are necessary for 2-controlled phase shift gates~\cite{yu2013five}, this decomposition is essentially optimal. 

For $k$-SAT with $k>3$, (ii) amounts to a $(k-1)$-controlled $\mathrm{C}^{k-1}\mathrm{Phase}(\gamma)$ gate. We can then iteratively decrease the number of controls by computing a pairwise AND into a $\ket{0}$-initialized ancilla qubit. This is illustrated for $4$-SAT in Figure~\ref{fig:PS-4SAT}:
Each such step reduces the number of controls by one, using 3 CNOTs and one ancilla.
As our base case we start with the CCPhase gate from Figure~\ref{fig:PS-clause}. 
Finally, all ancilla need to have an AND$^{\dagger}$ uncomputation. To further decrease the number of two-qubit gates, we make use of measurement-based uncomputations for $k$-SAT with $k>3$, which needs mid-circuit measurements, qubit reset, and classical feed-forward control. This trade-off is beneficially for ion-traps, which have high measurement fidelities. Measurement-based uncomputation amounts to 1 CZ gate per ancilla (which as an added benefit also allows us to move the last CNOT of our base case into a classically controlled single-qubit gate). 

Overall we need $k-3$ ancilla qubits and $4k-8$ two-qubit gates for any $k$-SAT clause with $k>3$ and 5 two-qubit gates per clause for $3$-SAT. 
Assuming a clause density of $4$ for $3$-SAT (justified later), we get a total of $20 \cdot n \cdot p$ two-qubit gates for a $p$-round QAOA circuit for a $3$-SAT problem on $n$ qubits, see Figures~\ref{fig:non_compiled_circuit} and~\ref{fig:compiled_circuit}.

\subsection{Quantum Hardware Implementation}
\label{section:methods_hardware_implementation}

The QAOA circuits are described using Qiskit \cite{Qiskit}, then transpiled using the Qiskit compiler in order to adapt the gateset to be as near to the quantum computer instruction set as possible. The final optimized compiled circuits that have had their instructions adapted are submitted to Quantinuum H1-1 via the Quantinuum API with the circuits described as Open QASM \cite{cross2017open} strings. The corresponding circuits are also submitted to IonQ Harmony via the Qiskit-IonQ plugin. The QAOA circuits are submitted to IonQ Aria 1 via the Python 3 Amazon Braket package. The QAOA circuits were executed on IonQ Forte by the internal IonQ team, with the QAOA circuits having been represented as OpenQASM files \cite{cross2017open}. Table \ref{table:hardware_summary} shows the qubit count on the $4$ quantum computers; the device qubit counts define the maximum $3$-SAT problem size that can be executed on that device. 

The circuits targeting Quantinuum H1-1 are compiled to have the gateset of \texttt{rzz, rz, ry, rx}. All four of these gates are single angle parameterized gates. The circuits targeting IonQ Harmony, IonQ Aria, and IonQ Forte are compiled to have the gateset of \texttt{cx, rz, ry, rx}. For circuit compilation to all $4$ quantum computers, the arbitrary (e.g. all-to-all) connectivity is used since all of the devices use the ion trap architecture which offers an all-to-all connectivity graph. An example $p=1$ $3$-SAT QAOA circuit that has been compiled to the \texttt{rzz, rz, ry, rx} gateset is shown in Figure \ref{fig:compiled_circuit}. 

\begin{table*}[ht!]
\begin{center}
\begin{tabular}{ |p{2.2cm}|p{1.3cm}|p{3.3cm}|p{3.3cm}|p{2.3cm}| }
 \hline
 QPU name & Number of Qubits & Avg. 1 qubit gate RB fidelity & Avg. 2 qubit gate RB fidelity & Avg. SPAM fidelity \\ 
 \hline
 \hline
 IonQ Harmony & 11 & 0.9958 & 0.9652 & 0.99752 \\ 
 \hline
 Quantinuum H1-1 & 20 & 0.99996 & 0.998 & 0.997 \\ 
 \hline
 IonQ Aria 1 & 25 & 0.981 & 0.9817 & 0.994 \\ 
 \hline
 IonQ Forte & 30 & 0.9998 & 0.99536 & 0.995 \\ 
 \hline
\end{tabular}
\end{center}
\caption{NISQ processor hardware summary used to execute the trained $3$-SAT QAOA circuits. }
\label{table:hardware_summary}
\end{table*}

\begin{figure*}[ht!]
    \centering
    \includegraphics[width=0.49\textwidth]{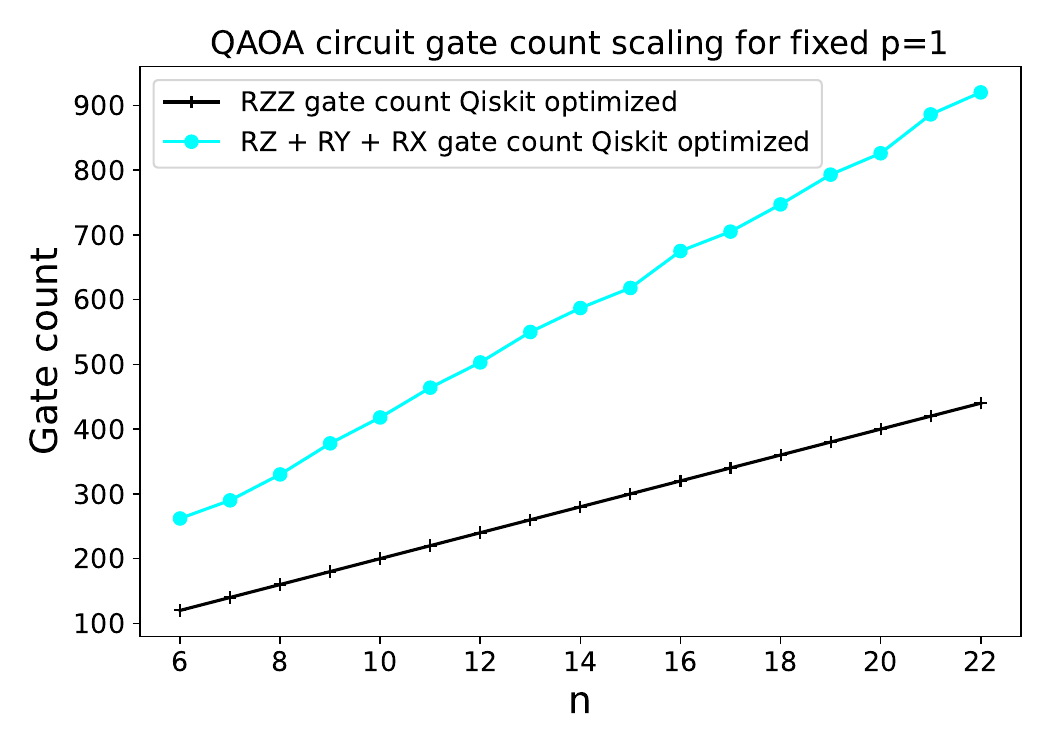}\hfill%
    \includegraphics[width=0.49\textwidth]{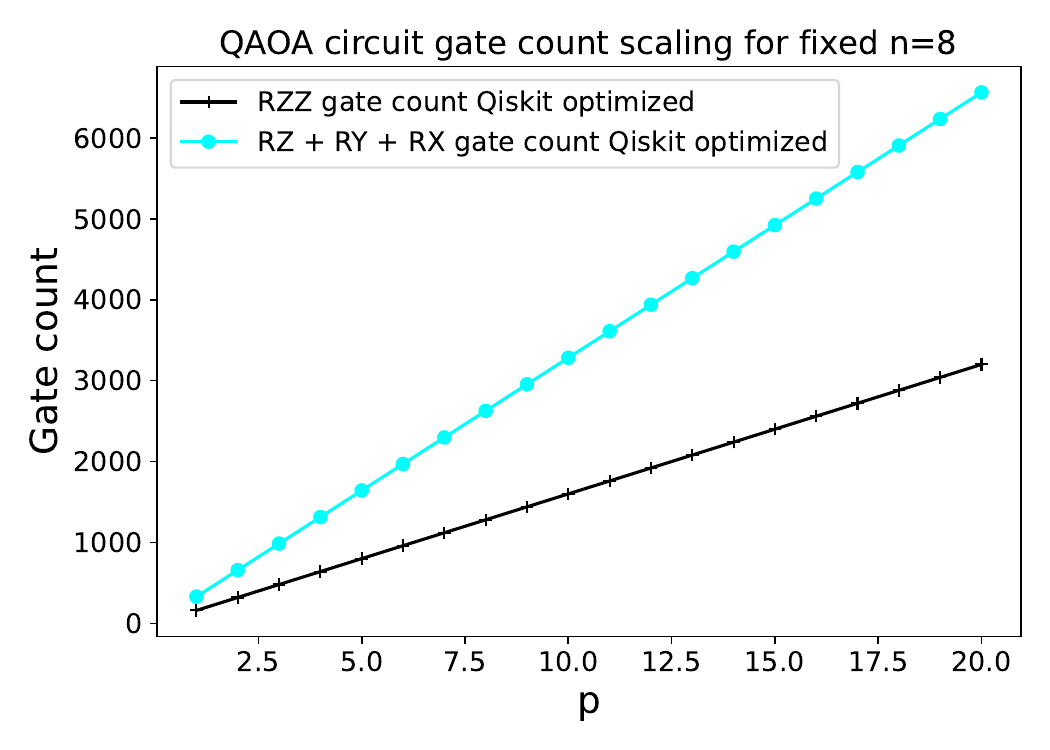}\\%
    \includegraphics[width=0.49\textwidth]{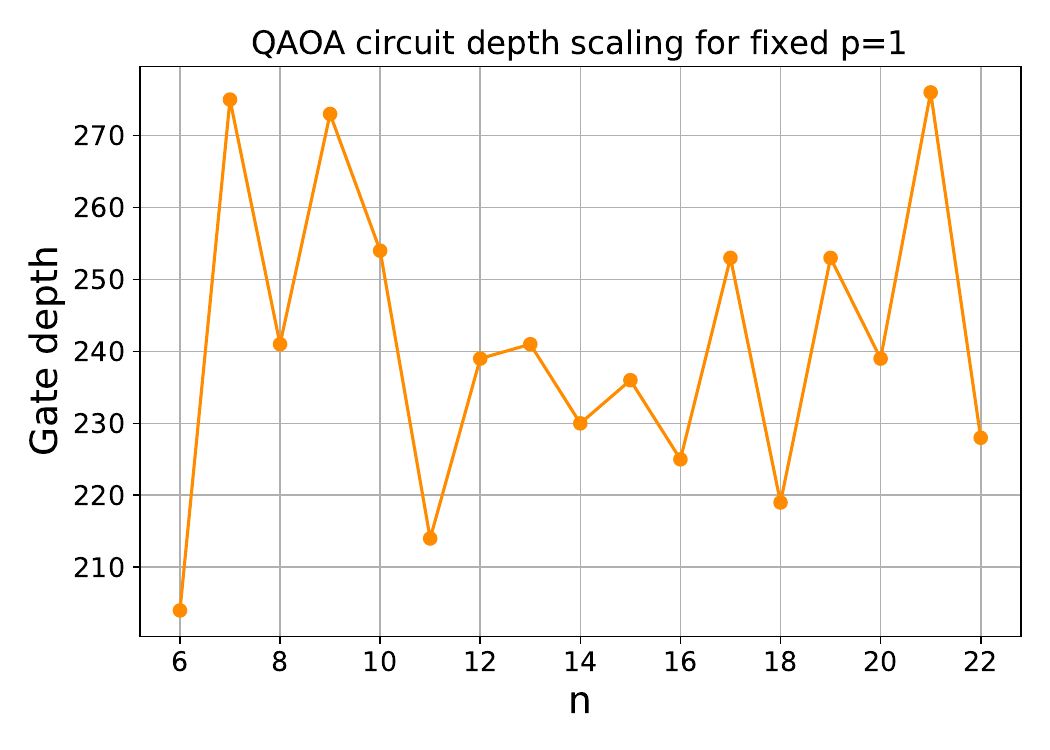}\hfill%
    \includegraphics[width=0.49\textwidth]{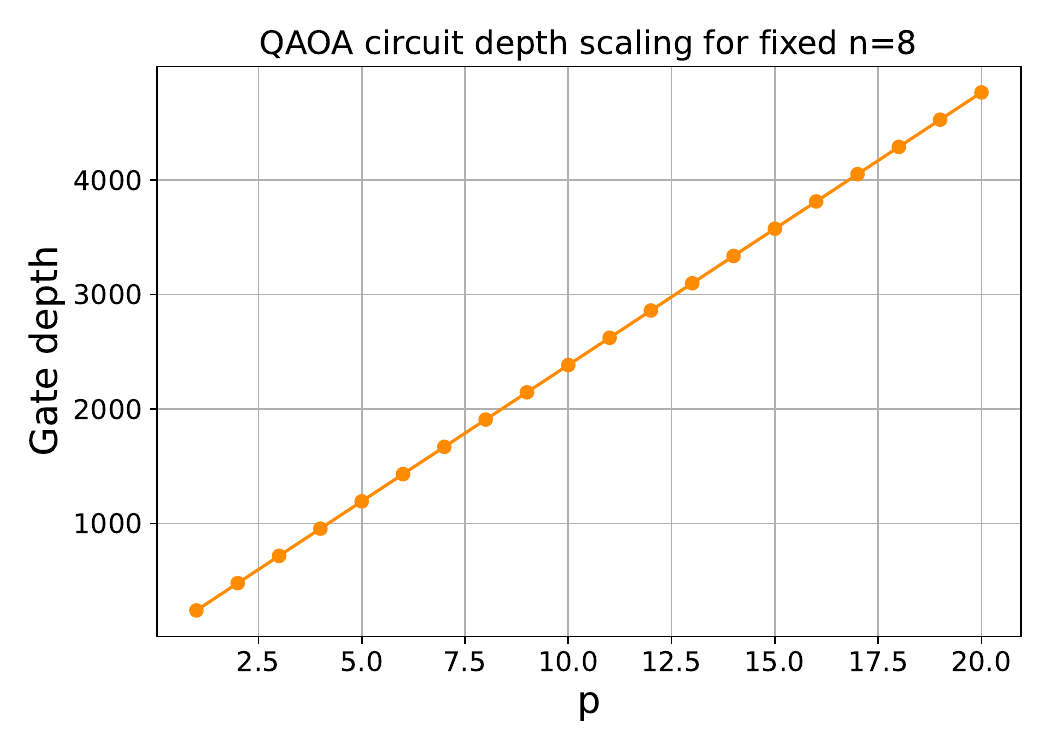}
    \caption{Top right: plotting single and two qubit gate counts in the generated QAOA circuits as function of increasing $p$ for a fixed $n=8$ $3$-SAT instance. The lower right plot shows the circuit depth (including both single and two qubit gates) for the same (increasing $p$) circuits as the top right plot. Top left: plotting the single and two qubit gate counts as a function of increasing the $3$-SAT problem size ($n$) for a fixed $p=1$. The lower left plot shows the circuit depth for the same increasing $n$ circuits from the top left plot. }
    \label{fig:gate_counts}
\end{figure*}

Figure \ref{fig:gate_counts} shows the scaling of the gate depth and single and two qubit gate counts for the (Qiskit compiled) QAOA circuits as a function of $p$ and $n$, for the gateset of \texttt{rzz, rz, ry, rx}. The gate depth plots in Figure \ref{fig:gate_counts} are computed using Qiskit \cite{Qiskit} (specifically \texttt{QuantumCircuit.depth}), and includes both single and two qubit gates. Interestingly, Figure \ref{fig:gate_counts} shows that as $n$ increases for a fixed $p$, the gate depth does not substantially increase for larger $n$, but rather seems oscillate within a few percentage points around an average. This is notable  because it implies that these QAOA circuits have fairly constant circuit depth per round even as we increase $n$; this is due to our choice of constant clause densities (in our case 4). 

For reference with respect to machine calibrations and features, all experiments were executed between April 2023 and June 2023. Quantinuum device backend parameters were set to default, which means that the pytket \cite{sivarajah2020t, cowtan_et_al} optimization level of $2$ (which is the highest possible optimization level in pytket) is applied server side. IonQ backend parameters are also all set to default for all of the IonQ device runs. 

All QAOA parameter combinations, e.g. for each $p$ and $n$, are sampled using at least $40$ shots (samples) on each quantum computer. This means that there will be shot noise present in the results.

For reference with respect to the expected device computation performance, we provide here the vendor-measured randomized benchmarking \cite{Magesan_2012, Harper_2019} calibration metrics for single qubit gates, two qubit gates, and SPAM (State Preparation and Measurement) error rates in Table \ref{table:hardware_summary}. Table Table \ref{table:hardware_summary} also gives the QPU name and the number of qubits available on that device. Because all $4$ of these quantum computers are trapped-ion based, their connectivity is all-to-all. The device calibration data for Quantinuum H1-1 was obtained from the device documentation during the timeframe in which the circuits were run. The device calibration data for IonQ Harmony and IonQ Aria 1 was obtained from the Amazon Braket device page within the time frame that the circuits were run. The device calibration data for IonQ Forte was measured using direct randomized benchmarking \cite{PhysRevLett.123.030503}, and described in ref. \cite{chen2023benchmarking}.

\subsection{Experimental Design}
\label{section:methods_experimental_design}
The essential question of ``how well does QAOA perform?'' is perhaps best thought of as two separate questions: ``assuming one has good (or optimal) angles, what kind of approximation ratio can QAOA give?'' and ''how difficult is it to find good angles?''. This work is particularly interested in the first question, with an emphasis on real-world performance on current NISQ devices. In this section we give details as to how we found the angles which are then tested on hardware, without significantly commenting on the difficulty (i.e. algorithmic scaling) of the angle finding. 

The combinatorial optimization problems that will be sampled on the NISQ computers are maximization random $3$-SAT instances, with a clause density of $4$. The selection of clause density is important because we need to have a balance between having too few clauses, which makes the optimization problem easy to solve, but each additional clause comes with substantial additional gate operations and higher circuit depths which means the computation will have more errors. We chose a clause density of $4$ based on ref. \cite{golden2023quantum}, which shows that a clause density of $4$ is right at the start of the plateau of the $3$-SAT problems being hard, which makes the problems hard in general to solve exactly and minimizes the number of circuit operations. 

For a given MAX 3-SAT instance and number of rounds $p$, we selected QAOA angles via a classical optimization algorithm with (exact, error free and shot noise free) QAOA simulations. In particular, our classical angle-finding scheme relies on combining a global optimization algorithm known as basin-hopping~\cite{basinhopping} with local gradient descent. We found that $10$ basin-hopping iterations produced high quality angles (though not necessarily optimal) up to $p=20$. This QAOA angle optimization code is implemented using the Julia \cite{bezanson2017julia} programming language. More discussion of our angle-finding approach, and comparison with other angle-finding schemes, can be found in refs.~\cite{golden2022evidence, golden2023quantum}.

For large problem sizes and high rounds, simulating the QAOA circuits is quite computationally intensive and requires HPC resources. Furthermore, moving forward it is clear that evaluating QAOA circuits via real hardware, rather than simulation, will be necessary for this style of angle-finding. However, we note that in general the scalability of real time angle-finding for a hybrid quantum-classical feedback loop is not clear for variational quantum algorithms, and in particular is expected to be quite hard especially with the presence of noise \cite{wang2021noise, Bittel_2021}. On the other hand, there is increasing evidence that classes of problems exhibit pronounced parameter concentration of the optimal or near optimal angles \cite{9605328, 9651381, Akshay_2021, brandao2018fixed, PhysRevA.104.052419, Farhi_2022, 8957201, pelofske2023quantum}, which in some cases could lead to scalable heuristics for good angle finding.

\begin{figure}[h!]
    \centering
    \includegraphics[width=\linewidth]{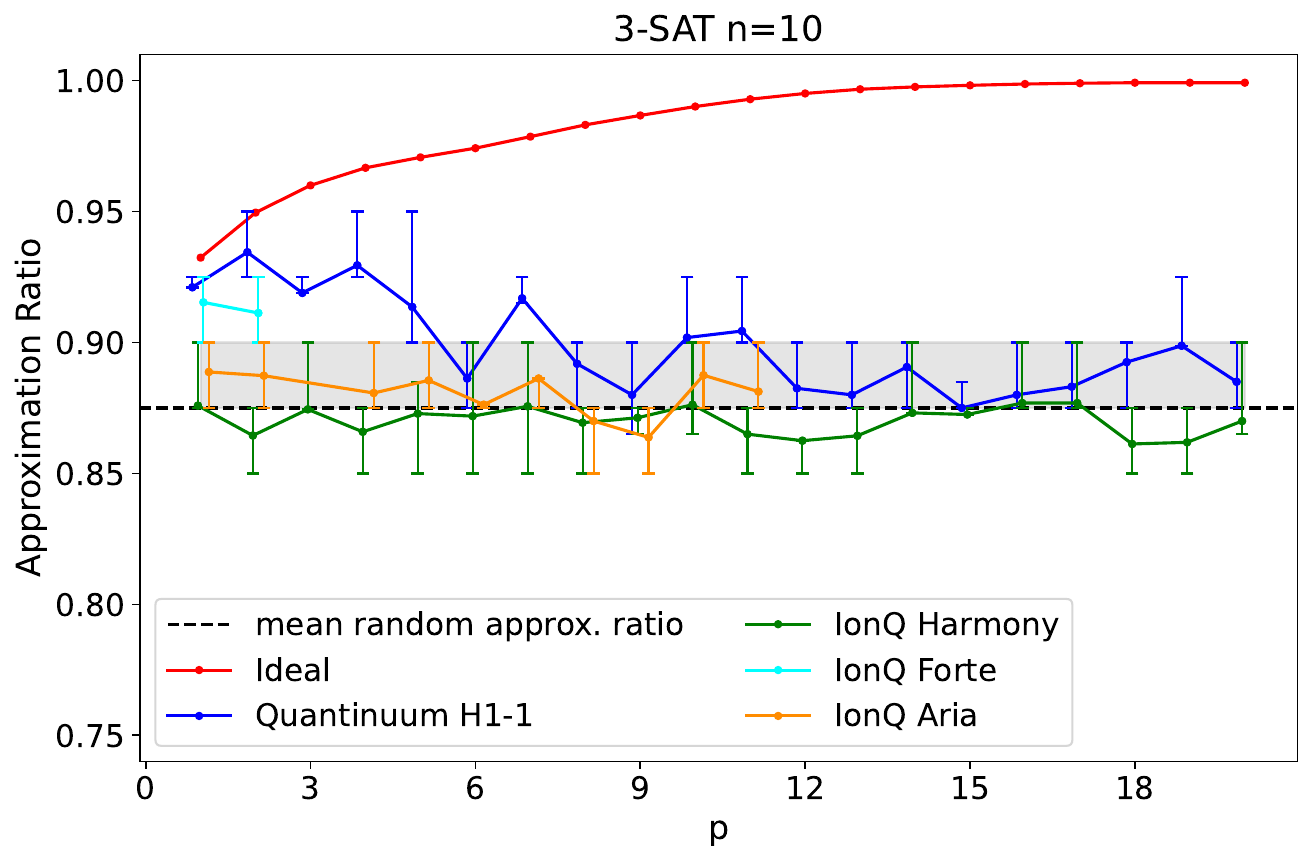}\\%
    \includegraphics[width=\linewidth]{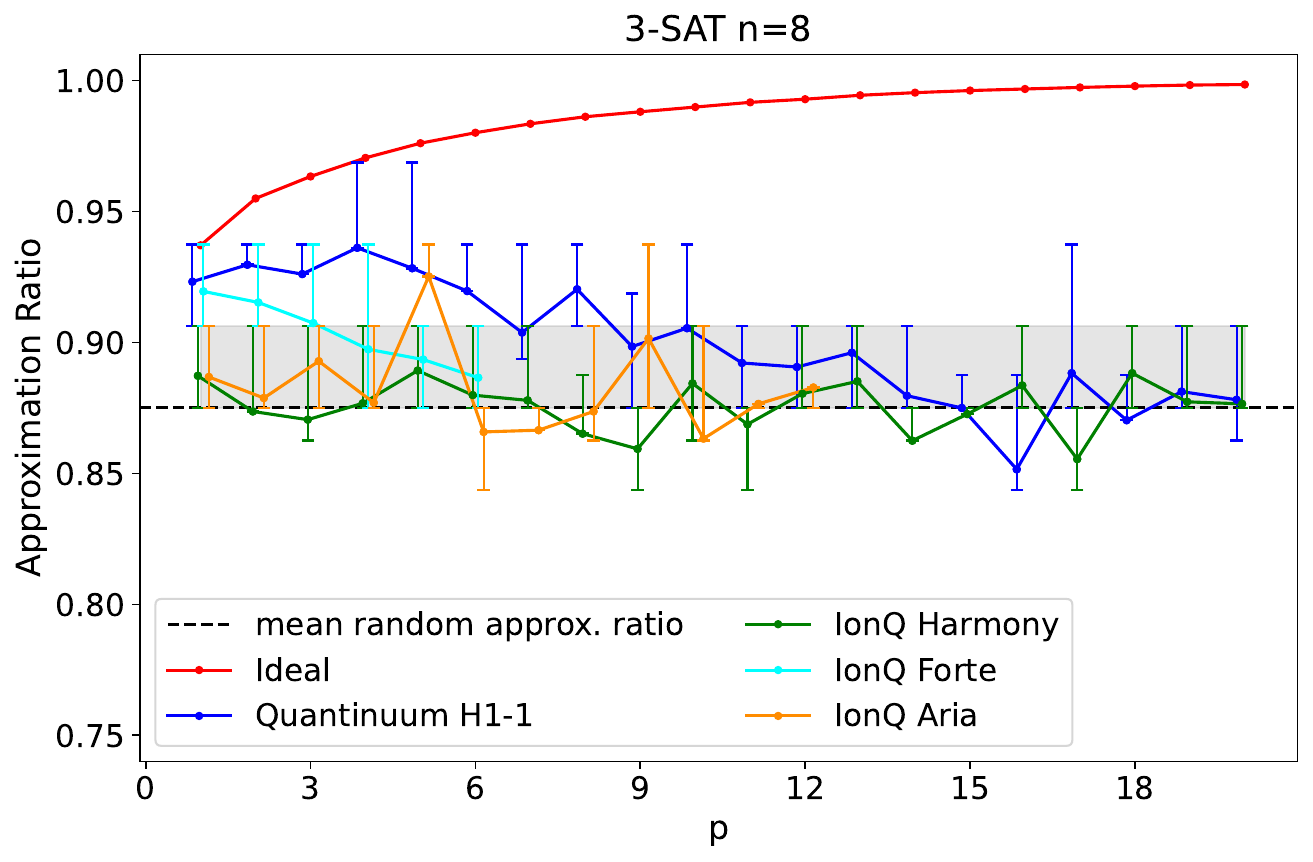}\\%
    \includegraphics[width=\linewidth]{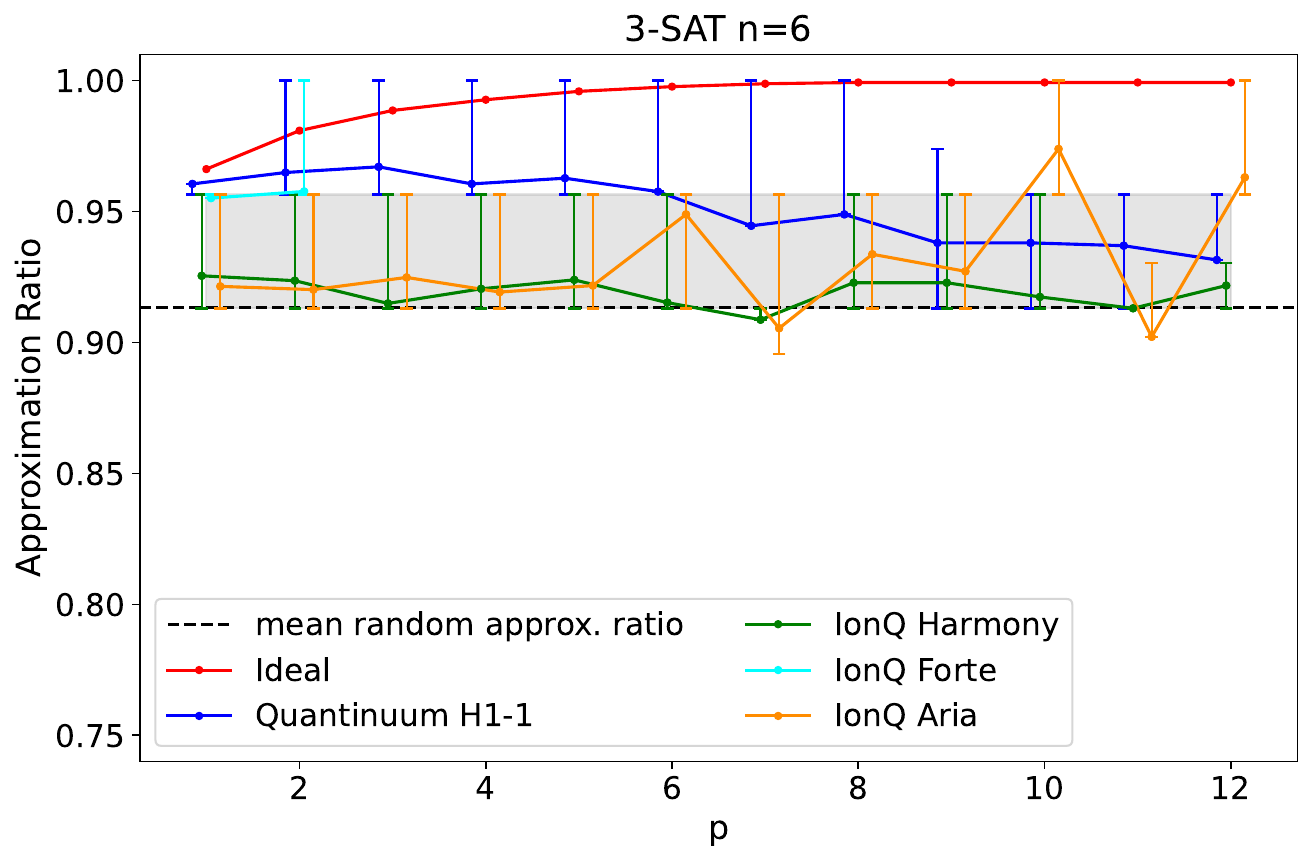}
    \caption{Approximation ratio (y-axis) vs the number of rounds (x-axis) when executing the $3$-SAT $n=10$, $n=8$, $n=6$ QAOA circuits on the $4$ trapped ion quantum computers. The mean approximation ratios are plotted as solid blue and green lines. The exact ideal approximation ratios for the classically computed angles (with no shot noise) are plotted in red. The 40th and 60th percentiles of random samples are shown by the shaded light grey region, the mean of the random sample distribution is plotted as the black dashed horizontal line. 
    }
    \label{fig:vary_rounds}
\end{figure}

\section{Results}
\label{section:results}

Figure \ref{fig:vary_rounds} shows the scaling of mean approximation ratio as a function of increasing the number of QAOA rounds $p$, for $n=10$, $n=8$, and $n=6$, respectively. The angles for each round were found using HPC classical computation in order to get very good, but not necessarily optimal, angles $\gamma$ and $\beta$. The classically learned angles allow the QAOA computation, without noise, to increase in approximation ratio monotonically, as shown by the red lines in Figure \ref{fig:vary_rounds}. The $n=6$ plot ends after $p=12$ because the ideal computation of the approximation ratio reaches above $0.999$, at which point the angle finding is terminated since the ideal computation has essentially plateaued and further rounds would not improve the computation substantially.

Because each additional round significantly adds to the total gate count, it is expected that the computation will accumulate errors at very high rounds and converge to random sampling. This is indeed what is observed in Figure \ref{fig:vary_rounds}, in particular for $n=10$ and $n=8$. Notably, rounds $p=1, 2, 3, 4$ for all three $3$-SAT problem instances on Quantinuum H1-1 are relatively stable in their mean approximation ratios, not immediately decreasing, but also not increasing substantially as is expected for the ideal QAOA computation. In Figure \ref{fig:vary_rounds} IonQ Harmony has a lower mean approximation ratio compared to Quantinuum H1-1, and in general there is not a clear signal that the sampled mean approximation ratio was measurably different than random sampling. The IonQ Aria samples in Figure \ref{fig:vary_rounds} perform similarly to IonQ Harmony. There is little IonQ Forte data in Figure \ref{fig:vary_rounds}, but the data that is present shows at low rounds ($p=1$ and $p=2$) the sample quality is comparable to Quantinuum H1-1. 

\begin{figure}[t!]
    \centering
    \includegraphics[width=\linewidth]{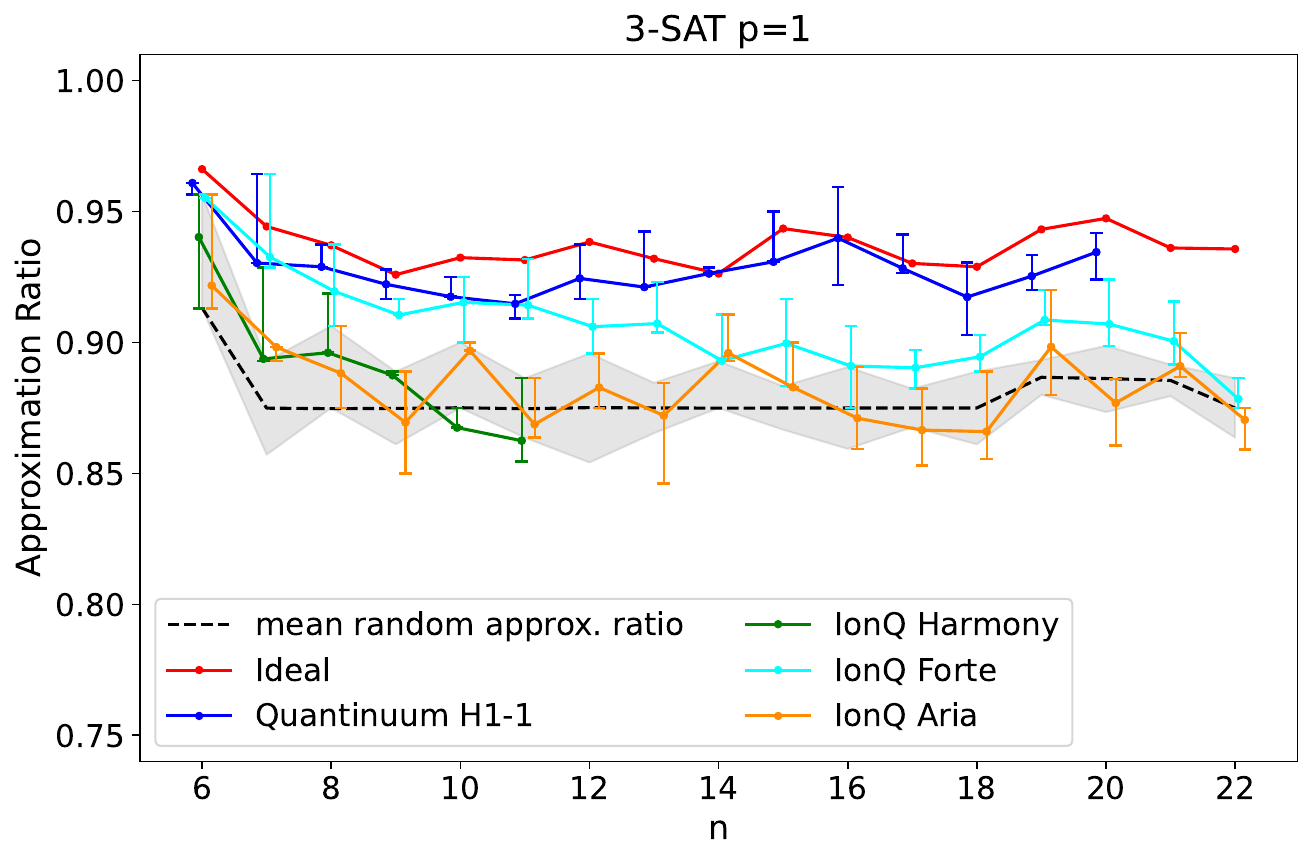}\\%
    \includegraphics[width=\linewidth]{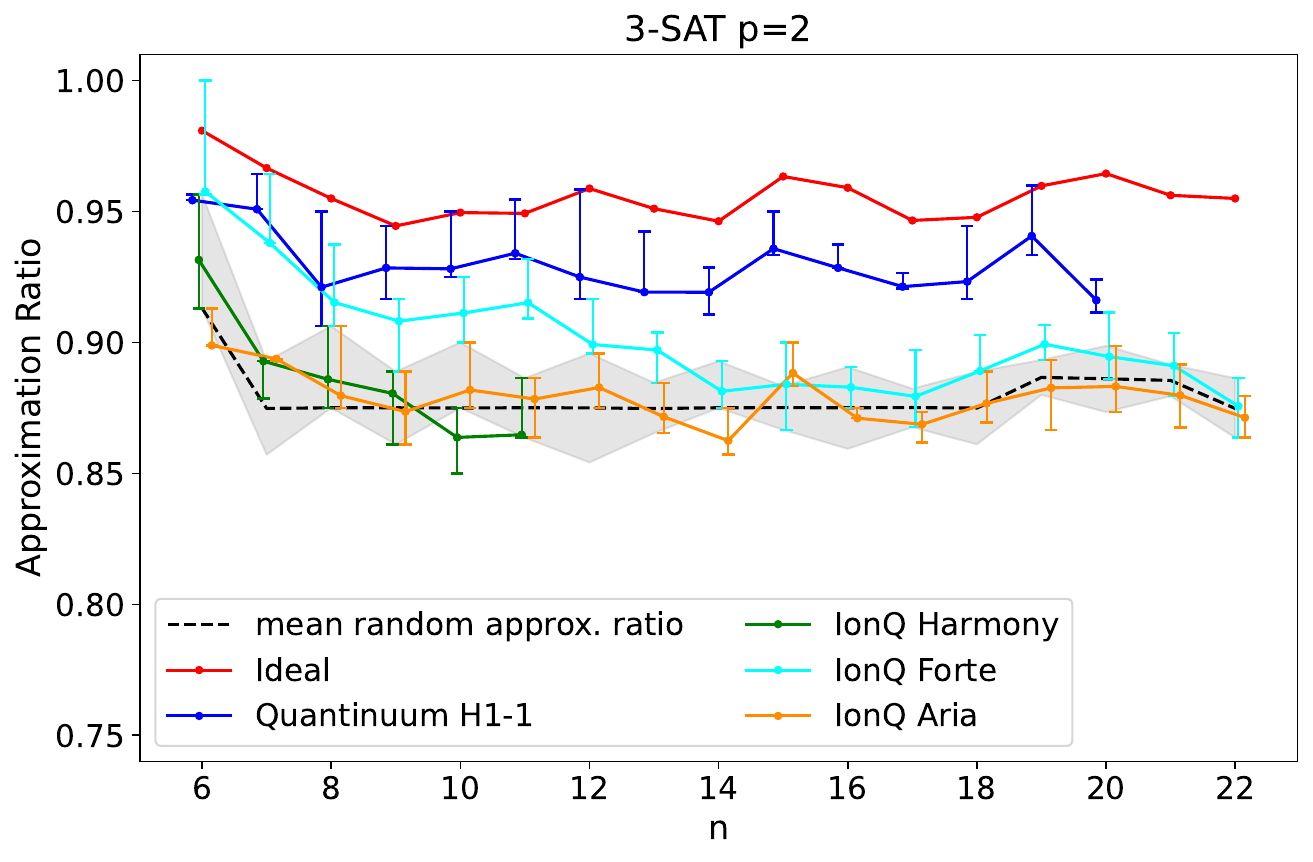}\\%
    \includegraphics[width=\linewidth]{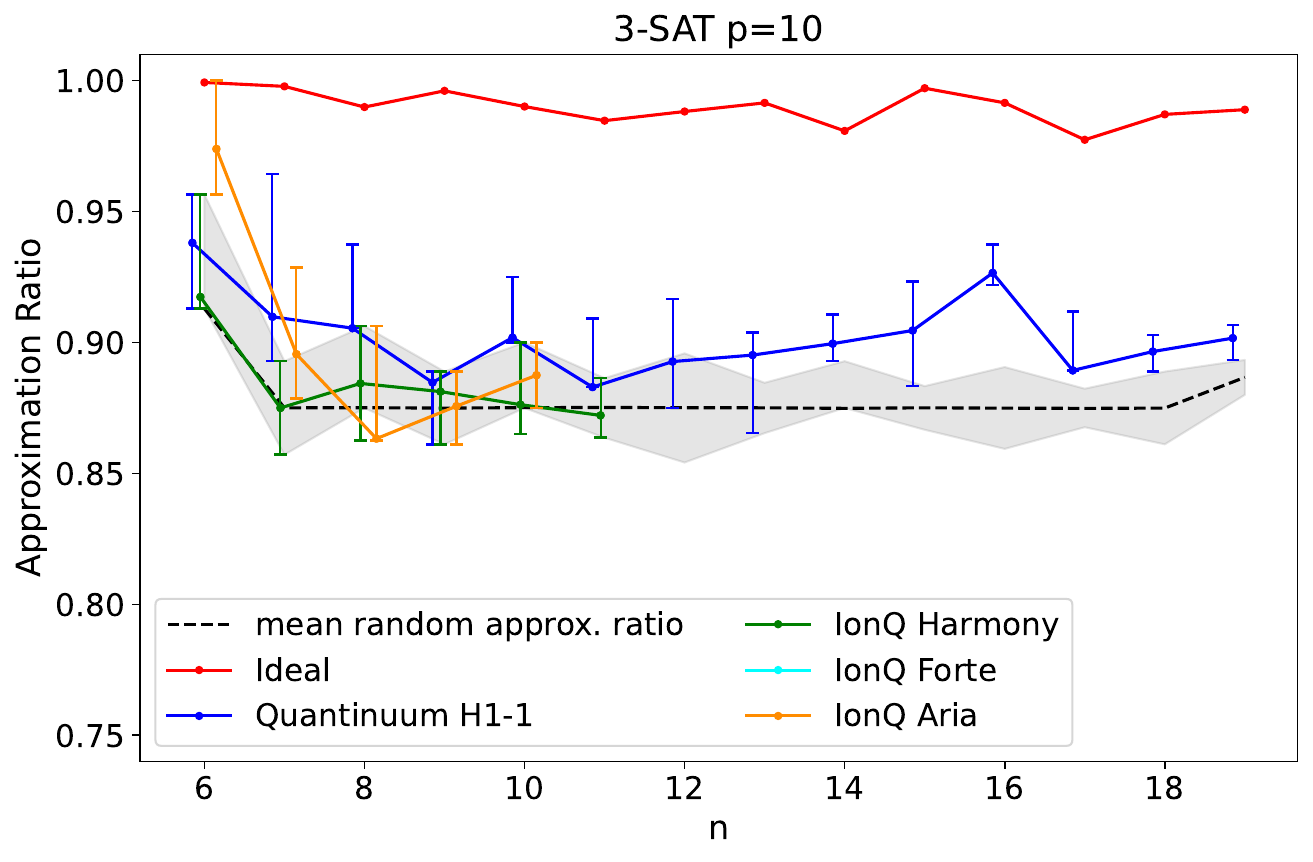}
    \caption{Approximation ratio (y-axis) vs the number of variables $n$ (x-axis) in the $3$-SAT problems for $p=1$, $p=2$, and $p=10$ QAOA circuits executed on the $4$ trapped ion quantum computers. The remaining figure details are the same as in Figure \ref{fig:vary_rounds}.}
    \label{fig:vary_size}
\end{figure}

Figure \ref{fig:vary_size} shows the scaling of mean approximation ratios as a function of increasing the $3$-SAT problem size, for a fixed number of rounds $p=1$, $p=2$, and $p=10$, respectively. Note that Figure \ref{fig:vary_size} goes up to $n=20$ for both $p=1$ and $p=2$, but only up to $n=19$ for $p=10$: the reason for the missing datapoint at $n=20$ for $10$ rounds is that the classical angle finding computation was too time intensive to compute in a reasonable amount of time. Interestingly, on Quantinuum H1-1 we observe that there remains a constant separation between the ideal approximation ratio and the NISQ computer sampled approximation ratio as a function of $n$ for both $p=1$ and $p=2$, suggesting that the computation does not significantly degrade as more qubits (e.g. variables) are added. Figure \ref{fig:gate_counts} shows that as $n$ increases the gate depth does not significantly change, which in part explains why there remains this constant approximation ratio gap where Quantinuum H1-1 still performs reasonably well up to $n=20$ for $p=1, 2$. Figure \ref{fig:vary_size} again shows that IonQ Harmony has a lower mean approximation ratio compared to the measured distribution from Quantinuum H1-1. The samples from IonQ Aria in Figure \ref{fig:vary_size} are comparable to IonQ Harmony. The data from IonQ Forte in Figure \ref{fig:vary_size} for $p=1$ and $p=2$ shows that the mean approximation ratios decrease to the noisy regime approximately at $n=14$, but at small $n$ IonQ Forte samples within the same regime of Quantinuum H1-1 (e.g. slightly below the ideal QAOA approximation ratios). 

Error bars are plotted in Figure \ref{fig:vary_rounds} and \ref{fig:vary_size} for all of the experiments. The error bars are asymmetric, and they are defined as the 40th and 60th percentiles of the sampled approximation ratio distributions so that the random sampling (light grey) shaded region can be compared against. The maximum ranges of the 40th and 60th percentile error bars are capped at the mean of the distribution for visually consistency of the plots. The error bars are visually slightly horizontally offset in order to make overlap of the distributions clear. 

The random sample distribution used in Figure \ref{fig:vary_rounds} and Figure \ref{fig:vary_size} (plotted as the light grey regions and the dashed black lines) are generated using $100,000$ (uniform) random variable assignment vectors, the approximation ratios of which are then evaluated. 

For both Figure \ref{fig:vary_rounds} and Figure \ref{fig:vary_size}, the number of shots used for IonQ Forte is $2500$ for the tested QAOA parameters and problem sizes. For the other three quantum computers the number of shots used was $40$ for each problem size and parameter combination, with the exception of the $n=6, 8, 10$ plots in Figure \ref{fig:vary_rounds} where a total of $140$ shots were used for $p=1, 2, 3, 4, 5$ (on IonQ Harmony, Quantinuum H1-1, and IonQ Aria) in order to obtain lower variance measurements at these low rounds in order to better determine whether the mean measured approximation ratio was clearly changing (increasing or decreasing) or staying constant. $140$ shots were also sampled for $n=8$, at $p=6, 7, 8$ rounds on IonQ Harmony, Quantinuum H1-1, and IonQ Aria. Examining the behavior of the solution quality at lower rounds is important because the accumulation of errors per round has a clear impact (performance degradation) on the actual execution of the algorithm of the algorithm even at low rounds.

The relative performance difference between Quantinuum H1-1, IonQ Harmony, IonQ Forte, and IonQ Aria are consistent with the vendor provided calibration metrics listed in Table \ref{table:hardware_summary}. Note that the incomplete data from IonQ Aria and IonQ Forte (incomplete in that there are datapoints missing that could have been executed given the device qubit count, see Table \ref{table:hardware_summary}) is in part due to a current backend error that prevents extremely high gate count circuits from being executed on those two devices. 

Let us return to our initial hypothesis of QAOA NISQ performance from Fig.~ \ref{fig:conceptual_diagram}. In both Figure \ref{fig:vary_rounds} and Figure \ref{fig:vary_size}, we have added light grey shaded regions that show the range between the 40th and 60th percentiles when randomly sampling the $3$-SAT instances to define $p_{\mathrm{noise}}$. Our conjectured peak point $p_{\mathrm{max}}$ is somewhat hard to identify due to shot noise, nevertheless it would be  $p_{\mathrm{max}} = 5, 4, 2$ for $n=6,8,10$, respectively, which overall is not as much of a dependence on instance size $n$. The round counts $p_{\mathrm{noise}}$ at which QAOA results become random are at $p_{\mathrm{noise}} = 6, 7, 6$ for $n=6,8,10$, respectively, as read off from Figure \ref{fig:vary_rounds} with little or no dependence on $n$ due the somewhat constant circuit depth. Therefore, a comparison against the random distribution becomes more meaningful when the problem size is significantly larger, e.g. the $n=15$ and greater distributions in Figure \ref{fig:vary_size}. As we would expect, the NISQ computers return mostly noise at $p=10$, however, when there is signal e.g., at $p=2$, this signal is almost independent of problem size, at least for Quantinuum H1-1.

The data generated in these experiments and the $3$-SAT problem instances are available as a public dataset \cite{elijah_pelofske_2023_8201693}.

\section{Discussion and Conclusion}
\label{section:conclusion}

Experimentally, these high round QAOA circuits are executed on the $4$ trapped ion quantum computers, IonQ Harmony, Quantinuum H1-1, IonQ Aria, and IonQ Forte. These QAOA circuits contain thousands of gate instructions, and constitute the largest experimental evaluation of QAOA circuits, with respect to gate depth and gate count, to date. These results show that very high round QAOA computations degrade into being random noise due the accumulation of errors, however we also observed that was not pronounced solution quality degradation for increasing the problem size $n$ for fixed small rounds $p=1$ and $p=2$. These experiments present a very strenuous evaluation of current quantum computing hardware, largely due to the random $3$-SAT instances having significantly higher QAOA gate depth per round (and total instruction count) compared to previous QAOA experiments on problems such as maximum cut \cite{https://doi.org/10.48550/arxiv.2303.02064} which can have much lower gate depths per QAOA round. 

Importantly, in these experiments the problem of angle finding on noisy quantum computers has been removed (e.g. offloaded to HPC simulations) with the intention of investigating the scaling of QAOA (with respect to problem size and number of rounds) on NISQ computers assuming reasonably good angles can be efficiently computed. However, future research could examine the angle learning aspect of these $k$-SAT circuits, along with associated questions such as what parameter concentrations can be observed, and whether those parameter concentrations offer a capability of transfer learning for $k$-SAT problems in QAOA as has been demonstrated for problems such as weighted maximum cut \cite{Shaydulin_2023}. Sampling tasks with QAOA are sensitive to shot noise, therefore future larger scale QAOA investigations should also also aim to increase the shot count significantly. Lastly, evaluating unexplored variants of QAOA mixers and state preparation algorithms applied to these $k$-SAT circuits would also be valuable future research. 

The experiments we present here did not use all existing trapped-ion quantum computers that are publicly accessible. Future research could utilize other existing trapped-ion quantum computers for even larger system sizes such as Quantinuum H2-1 \cite{moses2023race}. The primary difficulty for running this particular experiment type (e.g. heavily optimized QAOA circuits) on these larger systems, such as the 32 qubit Quantinuum H2-1 \cite{moses2023race}, is that the classical angle finding is approaching significant classical computational cost for an increasing number of QAOA rounds on high qubit count QAOA circuits. Therefore, future research on this topic will also need to include improved angle finding strategies, and optimized quantum circuit simulations for HPC systems.

\section{Acknowledgments}
\label{section:Acknowledgement}
All figures were generated using quantikz \cite{quantikz}, matplotlib \cite{Hunter:2007} and Qiskit \cite{Qiskit} in Python 3. 

We acknowledge the use of IonQ cloud and access to IonQ Harmony provided by IonQ. The authors thank the Quantinuum team, the IonQ team, and the Amazon Braket team for their technical support. 

This research used resources of the Oak Ridge Leadership Computing Facility, which is a DOE Office of Science User Facility supported under Contract DE-AC05-00OR22725. The Oak Ridge Leadership Computing Facility provided access to the Quantinuum H1-1 computer. This work was supported by the U.S. Department of Energy through the Los Alamos National Laboratory. Los Alamos National Laboratory is operated by Triad National Security, LLC, for the National Nuclear Security Administration of U.S. Department of Energy (Contract No. 89233218CNA000001). Research presented in this article was supported by the NNSA's Advanced Simulation and Computing Beyond Moore's Law Program at Los Alamos National Laboratory. This research used resources provided by the Darwin testbed at Los Alamos National Laboratory (LANL) which is funded by the Computational Systems and Software Environments subprogram of LANL's Advanced Simulation and Computing program (NNSA/DOE). We acknowledge the support of the LANL Information Science and Technology Institute. This work has been assigned LANL technical report number LA-UR-23-24237.


\bibliographystyle{plainurl}
\bibliography{references}

\end{document}